\documentclass[onecolumn,11pt,draftcls]{IEEEtran}
\fussy

\usepackage{amsmath,amssymb,verbatim,multirow,url,graphicx,epstopdf,cite}
\begin{document}
\title{Cyclotomic FFT of Length 2047 Based on a Novel 11-point Cyclic Convolution}
\author{Meghanad D. Wagh, Ning Chen, and Zhiyuan Yan}
\maketitle
\begin{abstract}
	In this manuscript, we propose a novel 11-point cyclic convolution algorithm based on alternate Fourier transform. With the proposed bilinear form, we construct a length-2047 cyclotomic FFT.
\end{abstract}

\section{Introduction}
Discrete Fourier transforms (DFTs) over finite fields have widespread applications in error correction coding~\cite{Blahut83}.
For Reed-Solomon (RS) codes, all syndrome-based bounded distance decoding methods involve DFTs over finite fields~\cite{Blahut83}: syndrome computation and the Chien search are both evaluations of polynomials and hence can be viewed as DFTs; inverse DFTs are used to recover transmitted codewords in transform-domain decoders.
Thus efficient DFT algorithms can be used to reduce the complexity of RS decoders.
For example, using the prime-factor fast Fourier transform (FFT) in~\cite{Truong06}, Truong {\it et al.} proposed~\cite{Truong06a} an inverse-free transform-domain RS decoder with substantially lower complexity than time-domain decoders; FFT techniques are used to compute syndromes for time-domain decoders in~\cite{Lin07}.

Cyclotomic FFT was proposed recently in~\cite{Trifonov03} and two variations were subsequently considered in~\cite{Costa04,Fedorenko06}.
Compared with other FFT techniques~\cite{Wang88,Truong06}, CFFTs in~\cite{Trifonov03,Costa04,Fedorenko06} achieve significantly lower multiplicative complexities, which makes them very attractive.
But their additive complexities (numbers of additions required) are very high if implemented directly.
A common subexpression elimination (CSE) algorithm was proposed to significantly reduce the additive complexities of CFFTs in~\cite{Chen08a}.
Along with those full CFFTs, reduced-complexity partial and dual partial CFFTs were used to design low complexity RS decoders in~\cite{Chen08b}.
The lengths of CFFTs in~\cite{Chen08a} are only up to 1023 while longer CFFTs are required to decode long RS codes.
To pursuit a length-2047 CFFT, 11-point cyclic convolution over characteristic-2 fields is necessary, which is not readily available in the literature.

In this manuscript, we first propose a novel 11-point cyclic convolution for characteristic-2 fields in Section~\ref{sec:conv}.
Based on this cyclic convolution, a length-2047 CFFT is presented in Section~\ref{sec:cfft}.
Using the same approach, CFFTs of any lengths that divide 2047 can also be constructed.

\section{11-Point Cyclic Convolution over Characteristic-2 Fields}\label{sec:conv}
We first derive a fast cyclic convolution of 11 points over the real field.
Denote the cyclic convolution of 11 point sequences $\boldsymbol{x}$ and $\boldsymbol{y}$ by the sequence $\boldsymbol{z}$.\footnote{In this manuscript, vectors and matrices are represented by boldface letters, and scalars by normal letters.}
The sequence $\boldsymbol{z}$ may be computed by Fourier transforming $\boldsymbol{x}$ and $\boldsymbol{y}$, multiplying the transforms point-by-point and finally, inverse Fourier transforming the product sequence.
%\subsection{Alternate Fourier Transform}
Let $\boldsymbol{X}$, $\boldsymbol{Y}$ and $\boldsymbol{Z}$ denote the Fourier transforms of $\boldsymbol{x}$, $\boldsymbol{y}$ and $\boldsymbol{z}$ respectively.
%We only need to compute $X_0$, $Y_0$, $\boldsymbol{X}'$ and $\boldsymbol{Y}'$.
As defined by the Fourier transform,
\begin{equation}
	X_0 = \sum_{i=0}^{10} x_i \quad Y_0 = \sum_{i=0}^{10} y_i.
	\label{X0}
\end{equation}
We express the rest components, $\boldsymbol{X}'$ and $\boldsymbol{Y}'$ (over reals) using the basis $\langle1, W, W^2, \dotsc, W^{9}\rangle$ where $W$ denotes the $11$th primitive root of unity.
This basis is sufficient because $W^{11} - 1 = 0$ yields $W^{10} = - 1 - W - W^2 - \dotsb - W^{9}$.
Thus, $\boldsymbol{X}' = \sum_{i=0}^{9} X'_i W^i$ and $\boldsymbol{Y}' = \sum_{i=0}^{9} Y'_i W^i$, in which
\begin{equation}
	X'_i = (x_i - x_{10}) \quad Y'_i = (y_i - y_{10}).
	\label{x-bar}
\end{equation}

We will call the vector $(X_0, X'_0, X'_1, \dotsc, X'_9)$ as the \emph{Alternate Fourier transform (AFT)} of sequence $\boldsymbol{x}$.
Note that AFT is simply the DFT components $X_0$ and $\boldsymbol{X}'$ in their special bases.
From \eqref{X0} and \eqref{x-bar} it is obvious that the AFT computation may be described as a multiplication with a $11 \times11$ matrix $\boldsymbol{B}$ with structure
$$
\boldsymbol{B} = \begin{bmatrix} 1 & 1 &  \dotso & 1 & 1\\
	& & & & -1\\
	& & \boldsymbol{I}_{10} & & -1\\
	& & & & \vdots\\
	& & & & -1
\end{bmatrix}$$
where $\boldsymbol{I}_{10}$ is a $10 \times 10$ identity matrix.
Alternately, given the AFT of $\boldsymbol{x}$, one can determine $\boldsymbol{x}$ by using matrix $\boldsymbol{B}^{-1}$ given by
\begin{equation}
	\boldsymbol{B}^{-1} = \frac{1}{11}\begin{bmatrix}
		1 & \boldsymbol{A}_1\\
		\boldsymbol{A}_2 & \boldsymbol{A}_3
	\end{bmatrix}
	\label{B_inverse}
\end{equation}
where length-10 row $\boldsymbol{A}_1 = (10, -1, -1,\dotsc, -1)$, length-10 column $\boldsymbol{A}_2 = (1, 1, \dotsc, 1)^\mathrm{T}$ and $10 \times 10$ submatrix $\boldsymbol{A}_3$ has 10 on the first upper diagonal and -1 everywhere else.

Now consider the product of $\boldsymbol{B}^{-1}$ and an AFT vector:
$$\boldsymbol{B}^{-1}\begin{bmatrix}U_0\\\boldsymbol{U}'\end{bmatrix} = \frac{1}{11} \begin{bmatrix}
		1 & \boldsymbol{A}_1\\
		\boldsymbol{A}_2 & \boldsymbol{A}_3
	\end{bmatrix}
	\begin{bmatrix}
		U_0\\ \boldsymbol{U}'
	\end{bmatrix} = \begin{bmatrix}V_0\\\boldsymbol{V}'
	\end{bmatrix}$$
where $U_0$, $\boldsymbol{U}'$, and $V_0$, $\boldsymbol{V}'$ are appropriate partitions of the AFT and the signal vectors.
Values of $V_0$ and $\boldsymbol{V}'$ can be computed as $V_0 = (1/11) U_0 + (1/11)\boldsymbol{A}_1 \boldsymbol{U}'$ and $\boldsymbol{V}' = (1/11) \boldsymbol{A}_2 U_0 + (1/11)\boldsymbol{A}_3 \boldsymbol{U}'$.
Note that $\boldsymbol{A}_1$ and $\boldsymbol{A}_3$ are related as $\boldsymbol{A}_1 = -(1, 1, \dotsc, 1)\,\boldsymbol{A}_3$.
This implies that the sum of the components of $(1/11) \boldsymbol{A}_3 \boldsymbol{U}'$ gives $-(1/11) \boldsymbol{A}_1 \boldsymbol{U}'$.
Furthermore, $\boldsymbol{A}_2$ contains only 1's.
Thus the computation of $V_0$ and $\boldsymbol{V}'$ reduces to 
\begin{equation}
	\label{V1_V2}
\begin{split}
	V_0 &= (1/11) U_0 - (1/11)\sum(\boldsymbol{A}_3 \boldsymbol{U}') \\ 
	\boldsymbol{V}' &= (1/11) [U_0, U_0, \dotsc, U_0]^\mathrm{T} + (1/11)\boldsymbol{A}_3 \boldsymbol{U}'.
\end{split}
\end{equation}
Relation \eqref{V1_V2} shows that the inverse of an AFT only needs an evaluation of $(1/11) \boldsymbol{A}_3 \boldsymbol{U}'$.

To compute cyclic convolution of $\boldsymbol{x}$ and $\boldsymbol{y}$, one should multiply the Fourier transforms of $\boldsymbol{x}$ and $\boldsymbol{y}$ and then take the inverse Fourier transform of the product.
We use AFT instead of classical Fourier transform.
Multiplying $X_0$ and $Y_0$ is simple, but since $\boldsymbol{X}'$ and $\boldsymbol{Y}'$ are expressed in a basis with 10 elements, their product may be difficult.
Similarly inverse AFT requires multiplication by matrix $\boldsymbol{A}_3$ which may be complicated.
However, we now show that both these two difficult computation stages are equivalent to only a Toeplitz product (i.e., product of a Toeplitz matrix and a vector)~\cite{Grenander58}.

The pointwise multiplication results are $Z_0 = X_0 Y_0$ and $\boldsymbol{Z}'$ defined as
%From the known representations of $\boldsymbol{X}'$, $\boldsymbol{Y}'$, and $\boldsymbol{Z}'$, we have
\begin{equation}
\biggl(\sum _{i=0}^{9} X'_i W^{i}\biggr)\biggl(\sum_{i=0}^{9} Y'_i W^i\biggr) = \sum _{i=0}^{9}Z'_iW^i.
\label{xy-product}
\end{equation}
Vector $\boldsymbol{Z}'$ can be computed through the matrix product $(Z'_0, Z'_1, \dotsc, Z'_9)^\mathrm{T} = \boldsymbol{M} (X'_0, X'_1, \dotsc, X'_9)^\mathrm{T}$ where the elements of matrix $\boldsymbol{M}$ are
\begin{equation}
M_{k, j} = Y'_{k-j} + Y'_{k-j+11} - Y'_{10-j}.
\label{M-ik} 
\end{equation}
Note that in \eqref{M-ik}, $Y'_{i}$ are considered as zero outside its valid range, i.e., $Y'_i = 0$ if $i < 0$ or $i > 9$.
The terms in \eqref{M-ik} are easy to deduce from \eqref{xy-product}.
Matrix element $M_{k,j}$ sums up those terms in $\boldsymbol{Y}'$ that after multiplication with $X'_j W^{j}$ result in $W^k$ terms.
For example, since product of $X'_j W^{j}$ and $Y'_{k-j} W^{k-j}$ results in $X'_jY'_{k-j} W^{k}$, we get the first term in \eqref{M-ik} as given.
Second term of \eqref{M-ik} can be similarly argued.
The third term is due to the product $(X'_j W^{j}) ( Y'_{10-j} W^{10-j} ) = X'_jY'_{10-j} W^{10} = - X'_jY'_{10-j} \sum_{i=0}^{9} W^{i}$.

Computing inverse DFT of $\boldsymbol{Z}$ requires one to multiply $\boldsymbol{A}_3$ and vector $(Z'_0, Z'_1, \dotsc, Z'_9)^T$ where $\boldsymbol{A}_{3}$ is the matrix defined in \eqref{B_inverse}.
Thus one has to compute $\boldsymbol{R} (X_1(0), X_1(1), \dotsc, X'_9)^T$ where the $10\times 10$ matrix $\boldsymbol{R} = (1/11)\boldsymbol{A}_3 \boldsymbol{M}$.

We now show by direct computation that $\boldsymbol{R}$ is a Toeplitz matrix.
From the structure of $\boldsymbol{A}_3$, we have
\begin{equation}
\label{R1}
\begin{split}
R_{i,j} &= \frac{1}{11}\sum_{k = 0, k \neq i+1}^{9} -M_{k, j} + \frac{10}{11} M_{i+1, j}\\
 &= -\frac{1}{11}\sum_{k=0}^{9} M_{k, j} + M_{i+1, j}.
\end{split}
\end{equation}
From \eqref{M-ik}, using the appropriate ranges for the three terms we
now get
\begin{equation}
\label{M-ki-sum}
\begin{split}
\sum_{k=0}^{9} M_{k, j} &= -\sum_{k=0}^{9} Y'_{10-j} + \sum_{k=0}^{j-2} Y'_{k-j+p} + \sum_{k=j}^{9} Y'_{k-j}\\
&= -10Y'_{10-j} + \sum_{s=11-j}^{9} Y'_s + \sum_{s=0}^{9-j} Y'_s\\
&= \sum_{s=0}^{9} Y'_s - 11 Y'_{10-j}
\end{split}
\end{equation}
Finally, combining \eqref{M-ik}, \eqref{R1} and \eqref{M-ki-sum} gives
\begin{equation}
	R_{i, j} = Y'_{i-j+1} + Y'_{i-j+12} - \frac{1}{11} \sum_{s=0}^{9} Y'_s.
\label{R2} 
\end{equation}
Since $R_{i,j}$ is a function of only $i-j$, $\boldsymbol{R}$ is a Toeplitz matrix. 
Thus $\boldsymbol{Z}'=\boldsymbol{R}\boldsymbol{X}'$ is computed as 
$$
\begin{bmatrix}
	Z'_0\\ Z'_1\\ \vdots\\ Z'_9
\end{bmatrix} =
\begin{bmatrix}
	Y'_1 & Y'_0 & 0 & Y'_9 & \dotso & Y'_3\\
	Y'_2 & Y'_1 & Y'_0 & 0 & \dotso & Y'_4\\
	\vdots & \vdots & \vdots & \vdots & \ddots & \vdots\\
	0 & Y'_9 & Y'_8 & \dotso & \dotso & Y'_1
\end{bmatrix}
\begin{bmatrix}
	X'_0\\ X'_1\\ \vdots\\ X'_9
\end{bmatrix} +
\begin{bmatrix}
	\sum_{i=0}^{9}X'_i\sum_{i=0}^{9}Y'_i\\ \sum_{i=0}^{9}X'_i\sum_{i=1}^{10}Y_i\\ \vdots\\ \sum_{i=0}^{9}X'_i\sum_{i=0}^{9}Y'_i
\end{bmatrix}.
$$

\begin{figure}[hbtp]
	\centering
	\includegraphics{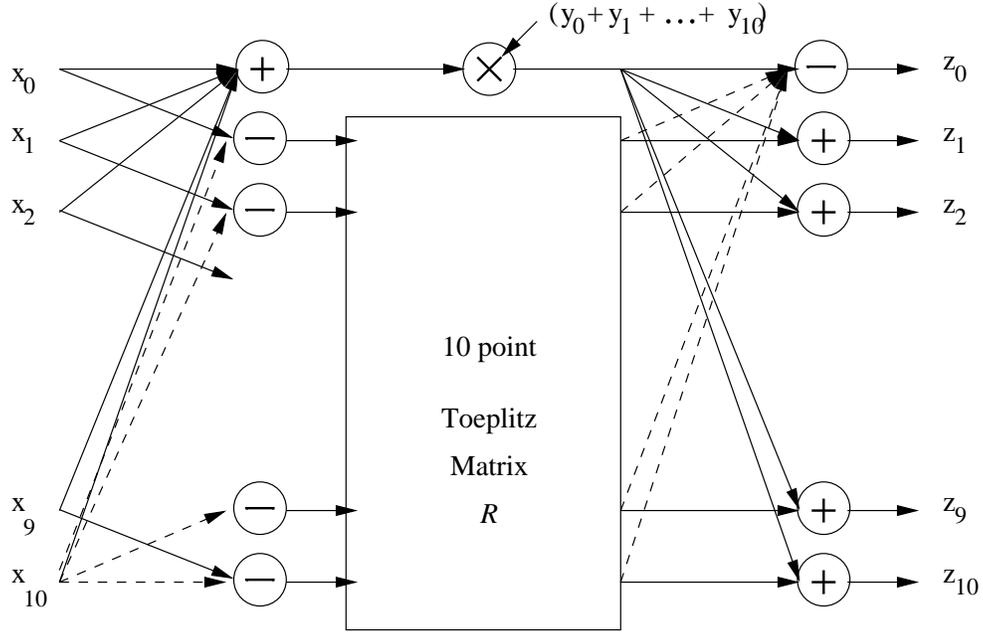}
	\caption{11-point cyclic convolution based on AFT}
	\label{fig:convo-11} 
\end{figure}

Recall that $Y'_i$ is assumed zero if its index is outside the valid range from 0 to 9.
Thus in \eqref{R2}, exactly one of the first two terms is valid for any combination of $i$ and $j$.
Fig.~\ref{fig:convo-11} illustrates the bilinear cyclic convolution algorithm of length 11 based on this discussion.  

%\subsection{Toeplitz Product of Length 10}
The multiplication of the $10\times 10$ Toeplitz matrix $\boldsymbol{R}$ with
a vector can be obtained using the Toeplitz product algorithms of lengths 2 and 5.
The matrix $\boldsymbol{R}$ can be split into four $5\times5$ submatrices and the vector $\boldsymbol{X}'$ can be split into two length-5 vectors.
By the definition of Toeplitz matrices, the Toeplitz product $\boldsymbol{R}\boldsymbol{X}'$ can be computed as
$$\boldsymbol{R}\boldsymbol{X}' = \begin{bmatrix}
	\boldsymbol{R}_0 & \boldsymbol{R}_1\\
	\boldsymbol{R}_2 & \boldsymbol{R}_0
\end{bmatrix}
\begin{bmatrix}
	\boldsymbol{X}'_0\\ \boldsymbol{X}'_1
\end{bmatrix} =
\begin{bmatrix}
	\boldsymbol{R}_0(\boldsymbol{X}'_0 + \boldsymbol{X}'_1) + (\boldsymbol{R}_1 - \boldsymbol{R}_0)\boldsymbol{X}'_1\\
	\boldsymbol{R}_0(\boldsymbol{X}'_0 + \boldsymbol{X}'_1) + (\boldsymbol{R}_2 - \boldsymbol{R}_0)\boldsymbol{X}'_0
\end{bmatrix}.$$

Although the cyclic convolution is derived over the real field, it can be easily converted to characteristic-2 fields.
Based on the method in~\cite{Blahut84}, we multiply both sides of all equations above by 11 modulus 2.
In the converted form, $\boldsymbol{X}=\boldsymbol{T}\boldsymbol{x}$, $\boldsymbol{Y}=\boldsymbol{T}\boldsymbol{y}$, and $\boldsymbol{z}=\boldsymbol{S}\boldsymbol{Z}$.
Thus we obtain 11-point cyclic convolution over characteristic-2 fields.
To find its bilinear form, we need the bilinear form of Toeplitz product of length 10. 
The bilinear form of length-5 Toeplitz product over characteristic-2 fields $\boldsymbol{v} = \boldsymbol{Q}^{(T5)}(\boldsymbol{R}^{(T5)}\boldsymbol{r} \cdot \boldsymbol{P}^{(T5)}\boldsymbol{u})$ is given in Appendix~\ref{sec:toeplitz5}, where $\cdot$ stands for pointwise multiplication.

%\subsection{Bilinear Form}
Based on the length-10 Toeplitz product, the bilinear form of 11-point cyclic convolution over $\mathrm{GF}(2^m)$ is given by
$$
\boldsymbol{z} = \boldsymbol{Q}^{(11)}(\boldsymbol{R}^{(11)}\boldsymbol{y}\cdot \boldsymbol{P}^{(11)}\boldsymbol{x}=\boldsymbol{S}
\begin{bmatrix}
	1 & \boldsymbol{0} & \boldsymbol{0} & \boldsymbol{0}\\
	\boldsymbol{0} & \boldsymbol{Q}^{(T5)} & \boldsymbol{Q}^{(T5)} & \boldsymbol{0}\\
	\boldsymbol{0} & \boldsymbol{Q}^{(T5)} & \boldsymbol{0} & \boldsymbol{Q}^{(T5)}
\end{bmatrix}
\left(\begin{bmatrix}
	1 & \boldsymbol{0} \\
	\boldsymbol{0} & \boldsymbol{R}^{(T5)}\boldsymbol{\Pi}_0\\
	\boldsymbol{0} & \boldsymbol{R}^{(T5)}\boldsymbol{\Pi}_1\\
	\boldsymbol{0} & \boldsymbol{R}^{(T5)}\boldsymbol{\Pi}_2
\end{bmatrix} \boldsymbol{T}\boldsymbol{y} \cdot
\begin{bmatrix}
	1 & \boldsymbol{0}\\
	\boldsymbol{0} & \boldsymbol{P}^{(T5)}\boldsymbol{\Pi}_3\\
	\boldsymbol{0} & \boldsymbol{P}^{(T5)}\boldsymbol{\Pi}_4\\
	\boldsymbol{0} & \boldsymbol{P}^{(T5)}\boldsymbol{\Pi}_5
\end{bmatrix} \boldsymbol{T}\boldsymbol{x}
\right).
$$
Details of matrices $\boldsymbol{S}, \boldsymbol{T}, \boldsymbol{\Pi}_0, \dotsc, \boldsymbol{\Pi}_5$ are given in Appendix~\ref{sec:conv11}.

The proposed length-11 cyclic convolution needs only 43 multiplications.
We compare it with cyclic convolutions of other lengths from~\cite{Blahut83,Wagh83,Trifonov} in Table~\ref{tab:conv}.

\begin{table}
	\centering
	\caption{Multiplicative Complexity of Cyclic Convolution}
	\begin{tabular}{|c|c|c|c|c|c|c|c|c|c|c|}
		\hline
		$n$ & 2 & 3 & 4 & 5 & 6 & 7 & 8 & 9 & 10 & 11\\
		\hline
		Mult. & 3 & 4 & 9 & 10 & 12 & 13 & 27 & 19 & 30 & 43\\
		\hline
	\end{tabular}
	\label{tab:conv}
\end{table}

\section{Cyclotomic FFT over $\mathrm{GF}(2^{11})$}\label{sec:cfft}
Based on the derived 11-point cyclic convolution over $\mathrm{GF}(2^m)$, we can construct a length-2047 cyclotomic FFT over $\mathrm{GF}(2^{11})$.
In this manuscript, we focus on direct CFFT as in~\cite{Trifonov03} since it was shown in~\cite{Chen08a} all variants of CFFTs have the same multiplicative complexity and they have the same additive complexity under direct implementation.

Given a primitive element $\alpha \in \mathrm{GF}(2^m)$, the DFT of a vector $\boldsymbol{f} = (f_0,f_1,\dotsc,f_{n-1})^T$ is defined as $\boldsymbol{F}\triangleq\bigl(f(\alpha^0), f(\alpha^1), \dotsc,
f(\alpha^{n-1})\bigr)^T$, where $f(x) \triangleq \sum_{i=0}^{n-1}f_i x^i
\in \mathrm{GF}(2^m)[x]$.

We choose the field generated by the polynomial $x^{11} + x^2 + 1$.
In this field, there are one size-1 coset and 186 size-11 cosets.
We permute the input $\boldsymbol{f}$ to $\boldsymbol{f}'$ such that $\boldsymbol{f}'=(f_0, \boldsymbol{f}_1, \boldsymbol{f}_2, \dotsc, \boldsymbol{f}_{186})$ and each size-11 vector $\boldsymbol{f}_i$ contains the components of $\boldsymbol{f}$ whose indices are in the same coset $(k_i, k_i2, \dotsc, k_i2^{m_i-1})\bmod 2047$ where $m_i \mid 11$ is the coset size.
Thus the polynomial $f(x)$ is divided into parts, each one is $L_i(x^{k_i})=\sum_{j=0}^{m_i-1}f_{k_i2^{j} \bmod 2047} (x^{k_i})^{2^{j}}$.
Hence $L_i(x)$'s are linearized polynomials.
Each element $\alpha^{k_i}$ can be decomposed with respect to a basis $\boldsymbol{\beta}_i=(\beta_{i,0}, \beta_{i,1}, \dotsc, \beta_{i,m_i-1})$ such that $\alpha^{jk_i} = \sum_{s=0}^{10} a_{i,j,s}\beta_{i,s}, a_{i,j,s} \in \mathrm{GF}(2)$.
So each component of DFT is factored into
$$f(\alpha^j)=\sum_{i=0}^{186}L_i(\alpha^{jk_i})=\sum_{i=0}^{186}\sum_{s=0}^{m_i}a_{i,j,s}L_i(\beta_{i,s})= \sum_{i=0}^{186}\sum_{s=0}^{10} a_{i,j,s} \bigl( \sum_{p=0}^{10} \beta_{i,s}^{2^p} f_{k_i2^p}\bigr).$$
In matrix form, it is $\boldsymbol{F}=\boldsymbol{a}\boldsymbol{L}\boldsymbol{f}$, in which $\boldsymbol{L}$ is a block diagonal matrix with each diagonal block being
$$ \boldsymbol{L}_i = 
\begin{bmatrix}
	\beta_{i,0} & \beta_{i,0}^2 & \dotso & \beta_{i,0}^{2^m_i-1}\\
	\beta_{i,1} & \beta_{i,1}^2 & \dotso & \beta_{i,1}^{2^m_i-1}\\
	\vdots & \vdots & \ddots & \vdots\\
	\beta_{i,m_i-1} & \beta_{i,m_i-1}^2 & \dotso & \beta_{i,m_i-1}^{2^m_i-1}
\end{bmatrix}.$$
Using a normal basis as $\boldsymbol{\beta}_i$, the matrix $\boldsymbol{L}_i$ becomes a cyclic matrix and $\boldsymbol{L}_i \boldsymbol{f}_i$ becomes a size-$m_i$ cyclic convolution. 
For length-2047 CFFT, $m_i$ is 1 or 11.
Thus we obtain a length-2047 CFFT using the bilinear form of 11-point cyclic convolution as
$$\boldsymbol{F}= \boldsymbol{a}
	\begin{bmatrix} 1\\
		& \boldsymbol{Q}^{(11)}\\
		& & \ddots\\
		& & & \boldsymbol{Q}^{(11)}
	\end{bmatrix}\left(
	\begin{bmatrix} 1\\
		& \boldsymbol{R}^{(11)}\\
		& & \ddots\\
		& & & \boldsymbol{R}^{(11)}
	\end{bmatrix}
	\begin{bmatrix} 1\\
		\boldsymbol{\beta}_1\\
		\vdots\\
		\boldsymbol{\beta}_{186}
	\end{bmatrix}
	\cdot
	\begin{bmatrix} 1\\
		& \boldsymbol{P}^{(11)}\\
		& & \ddots\\
		& & & \boldsymbol{P}^{(11)}
	\end{bmatrix}
	\begin{bmatrix} f_0\\
		\boldsymbol{f}_1\\
		\vdots\\
		\boldsymbol{f}_{186}
	\end{bmatrix}\right).
$$
It requires 7812 multiplications to compute the constructed length-2047 CFFT.
Under direct implementation, it requires $2154428$ additions.
With incomplete optimization using the CSE algorithm~\cite{Chen08a}, its additive complexity can be reduced to $529720$.
We compare its complexity with those of shorter CFFTs in Table~\ref{tab:cfft}. In Table~\ref{tab:cfft}, our numbers of additions for CFFTs of lengths $7, 15, \dotsb, 1023$ are reproduced from~\cite{Chen08a}.
\begin{table}[htbp]
	\centering
	\caption{Complexity of Full Cyclotomic FFT}
	\label{tab:cfft}
	\begin{tabular}{|c|c|c|c|c|c|c|c|}
			\hline
			\multirow{2}{*}{$n$} & \multirow{2}{*}{Mult.} & \multicolumn{3}{c|}{Additions}\\
			\cline{3-5}
			& & Ours & \cite{Trifonov03} & Direct\\
			\hline
			7 & 6 & 24 & 25 & 34\\
			\hline
			15 & 16 & 74 & 77 & 154\\
			\hline
			31 & 54 & 299 & 315 & 570\\
			\hline
			63 & 97 & 759 & 805 & 2527\\
			\hline
			127 & 216 & 2576 & 2780 & 9684\\
			\hline
			255 & 586 & 6736 & 7919 & 37279\\
			\hline
			511 & 1014 & 23130 & 26643 & 141710\\
			\hline
			1023 & 2827 & 75360 & - & 536093\\
			\hline
			2047 & 7812 & 529720 & - & 2154428\\
			\hline
	\end{tabular}
\end{table}

\appendices
\section{Toeplitz Product of Length 5}\label{sec:toeplitz5}
Toeplitz product of length 5 as
$$\begin{bmatrix}
	v_0\\ v_1\\ v_2\\ v_3\\ v_4
\end{bmatrix}
=
\begin{bmatrix}
	r_4 & r_5 & r_6 & r_7 & r_8\\
	r_3 & r_4 & r_5 & r_6 & r_7\\
	r_2 & r_3 & r_4 & r_5 & r_6\\
	r_1 & r_2 & r_3 & r_4 & r_5\\
	r_0 & r_1 & r_2 & r_3 & r_4
\end{bmatrix}
\begin{bmatrix}
	u_0\\ u_1\\ u_2\\ u_3\\ u_4
\end{bmatrix}$$
can be done in bilinear form as
$\boldsymbol{v} = \boldsymbol{Q}^{(T5)}(\boldsymbol{R}^{(T5)}\boldsymbol{r} \cdot \boldsymbol{P}^{(T5)}\boldsymbol{u})$.

\begin{align*}
	\boldsymbol{R}^{(T5)} = \begin{bmatrix}
	1 & 1 & 1 & 1 & 1 & 0 & 0 & 0 & 0\\
	0 & 1 & 1 & 1 & 1 & 1 & 0 & 0 & 0\\
	0 & 0 & 1 & 1 & 1 & 1 & 1 & 0 & 0\\
	0 & 0 & 0 & 1 & 1 & 1 & 1 & 1 & 0\\
	0 & 0 & 0 & 0 & 1 & 1 & 1 & 1 & 1\\
	0 & 1 & 0 & 0 & 1 & 0 & 0 & 0 & 0\\
	0 & 0 & 1 & 0 & 0 & 0 & 0 & 0 & 0\\
	0 & 0 & 0 & 1 & 1 & 0 & 0 & 0 & 0\\
	0 & 0 & 0 & 1 & 0 & 0 & 0 & 0 & 0\\
	0 & 0 & 0 & 0 & 1 & 1 & 0 & 0 & 0\\
	0 & 0 & 0 & 0 & 0 & 1 & 0 & 0 & 0\\
	0 & 0 & 0 & 0 & 0 & 0 & 1 & 0 & 0\\
	0 & 0 & 0 & 0 & 1 & 0 & 0 & 1 & 0\\
	0 & 0 & 0 & 0 & 1 & 0 & 0 & 0 & 0
	\end{bmatrix} \quad
	\boldsymbol{P}^{(T5)} = \begin{bmatrix}
	1 & 0 & 0 & 0 & 0\\
	0 & 1 & 0 & 0 & 0\\
	0 & 0 & 1 & 0 & 0\\
	0 & 0 & 0 & 1 & 0\\
	0 & 0 & 0 & 0 & 1\\
	1 & 1 & 0 & 0 & 0\\
	1 & 0 & 1 & 0 & 0\\
	1 & 0 & 0 & 1 & 0\\
	0 & 1 & 1 & 0 & 0\\
	0 & 1 & 0 & 0 & 1\\
	0 & 0 & 1 & 1 & 0\\
	0 & 0 & 1 & 0 & 1\\
	0 & 0 & 0 & 1 & 1\\
	1 & 1 & 0 & 1 & 1
	\end{bmatrix}\\
	\boldsymbol{Q}^{(T5)} = \begin{bmatrix}
	0 & 0 & 0 & 0 & 1 & 0 & 0 & 0 & 0 & 1 & 0 & 1 & 1 & 1\\
	0 & 0 & 0 & 1 & 0 & 0 & 0 & 1 & 0 & 0 & 1 & 0 & 1 & 1\\
	0 & 0 & 1 & 0 & 0 & 0 & 1 & 0 & 1 & 0 & 1 & 1 & 0 & 0\\
	0 & 1 & 0 & 0 & 0 & 1 & 0 & 0 & 1 & 1 & 0 & 0 & 0 & 1\\
	1 & 0 & 0 & 0 & 0 & 1 & 1 & 1 & 0 & 0 & 0 & 0 & 0 & 1
	\end{bmatrix}.
\end{align*}

\section{Bilinear Form of 11-point Cyclic Convolution over Characteristic-2 Fields}\label{sec:conv11}
$$
\boldsymbol{z} = \boldsymbol{S}
\begin{bmatrix}
	1\\
	& \boldsymbol{Q}^{(T5)} & \boldsymbol{Q}^{(T5)} & \boldsymbol{0}\\
	& \boldsymbol{Q}^{(T5)} & \boldsymbol{0} & \boldsymbol{Q}^{(T5)}
\end{bmatrix}
\left(\begin{bmatrix}
	1\\
	& \boldsymbol{R}^{(T5)}\boldsymbol{\Pi}_0\\
	& \boldsymbol{R}^{(T5)}\boldsymbol{\Pi}_1\\
	& \boldsymbol{R}^{(T5)}\boldsymbol{\Pi}_2
\end{bmatrix} \boldsymbol{T}\boldsymbol{y} \cdot
\begin{bmatrix}
	1\\
	& \boldsymbol{P}^{(T5)}\boldsymbol{\Pi}_3\\
	& \boldsymbol{P}^{(T5)}\boldsymbol{\Pi}_4\\
	& \boldsymbol{P}^{(T5)}\boldsymbol{\Pi}_5
\end{bmatrix} \boldsymbol{T}\boldsymbol{x}
\right).
$$
$$
\boldsymbol{T}=
\begin{bmatrix}
	1 & 1 & 1 & 1 & 1 & 1 & 1 & 1 & 1 & 1\\
	1 & 0 & 0 & 0 & 0 & 0 & 0 & 0 & 0 & 1\\
	0 & 1 & 0 & 0 & 0 & 0 & 0 & 0 & 0 & 1\\
	0 & 0 & 1 & 0 & 0 & 0 & 0 & 0 & 0 & 1\\
	0 & 0 & 0 & 1 & 0 & 0 & 0 & 0 & 0 & 1\\
	0 & 0 & 0 & 0 & 1 & 0 & 0 & 0 & 0 & 1\\
	0 & 0 & 0 & 0 & 0 & 1 & 0 & 0 & 0 & 1\\
	0 & 0 & 0 & 0 & 0 & 0 & 1 & 0 & 0 & 1\\
	0 & 0 & 0 & 0 & 0 & 0 & 0 & 1 & 0 & 1\\
	0 & 0 & 0 & 0 & 0 & 0 & 0 & 0 & 1 & 1
\end{bmatrix} \quad
\boldsymbol{S} = 
\begin{bmatrix}
	1 & 1 & 1 & 1 & 1 & 1 & 1 & 1 & 1 & 1 & 1\\
	1 & 1 & 0 & 0 & 0 & 0 & 0 & 0 & 0 & 0 & 0\\
	1 & 0 & 1 & 0 & 0 & 0 & 0 & 0 & 0 & 0 & 0\\
	1 & 0 & 0 & 1 & 0 & 0 & 0 & 0 & 0 & 0 & 0\\
	1 & 0 & 0 & 0 & 1 & 0 & 0 & 0 & 0 & 0 & 0\\
	1 & 0 & 0 & 0 & 0 & 1 & 0 & 0 & 0 & 0 & 0\\
	1 & 0 & 0 & 0 & 0 & 0 & 1 & 0 & 0 & 0 & 0\\
	1 & 0 & 0 & 0 & 0 & 0 & 0 & 1 & 0 & 0 & 0\\
	1 & 0 & 0 & 0 & 0 & 0 & 0 & 0 & 1 & 0 & 0\\
	1 & 0 & 0 & 0 & 0 & 0 & 0 & 0 & 0 & 1 & 0\\
	1 & 0 & 0 & 0 & 0 & 0 & 0 & 0 & 0 & 0 & 1
\end{bmatrix}$$
$$\boldsymbol{\Pi}_0 = 
\begin{bmatrix}
	1 & 1 & 1 & 1 & 1 & 0 & 1 & 1 & 1 & 1\\
	1 & 1 & 1 & 1 & 0 & 1 & 1 & 1 & 1 & 1\\
	1 & 1 & 1 & 0 & 1 & 1 & 1 & 1 & 1 & 1\\
	1 & 1 & 0 & 1 & 1 & 1 & 1 & 1 & 1 & 1\\
	1 & 0 & 1 & 1 & 1 & 1 & 1 & 1 & 1 & 1\\
	0 & 1 & 1 & 1 & 1 & 1 & 1 & 1 & 1 & 1\\
	1 & 1 & 1 & 1 & 1 & 1 & 1 & 1 & 1 & 1\\
	1 & 1 & 1 & 1 & 1 & 1 & 1 & 1 & 1 & 0\\
	1 & 1 & 1 & 1 & 1 & 1 & 1 & 1 & 0 & 1
\end{bmatrix} \quad
\boldsymbol{\Pi}_3 =
\begin{bmatrix}
	1 & 0 & 0 & 0 & 0 & 1 & 0 & 0 & 0 & 0\\
	0 & 1 & 0 & 0 & 0 & 0 & 1 & 0 & 0 & 0\\
	0 & 0 & 1 & 0 & 0 & 0 & 0 & 1 & 0 & 0\\
	0 & 0 & 0 & 1 & 0 & 0 & 0 & 0 & 1 & 0\\
	0 & 0 & 0 & 0 & 1 & 0 & 0 & 0 & 0 & 1
\end{bmatrix}
$$

$$\boldsymbol{\Pi}_1 = 
\begin{bmatrix}
	1 & 0 & 0 & 0 & 0 & 1 & 0 & 0 & 0 & 0\\
	0 & 0 & 0 & 0 & 1 & 0 & 0 & 0 & 0 & 0\\
	0 & 0 & 0 & 1 & 0 & 0 & 0 & 0 & 0 & 1\\
	0 & 0 & 1 & 0 & 0 & 0 & 0 & 0 & 1 & 0\\
	0 & 1 & 0 & 0 & 0 & 0 & 0 & 1 & 0 & 0\\
	1 & 0 & 0 & 0 & 0 & 0 & 1 & 0 & 0 & 0\\
	0 & 0 & 0 & 0 & 0 & 1 & 0 & 0 & 0 & 0\\
	0 & 0 & 0 & 0 & 1 & 0 & 0 & 0 & 0 & 1\\
	0 & 0 & 0 & 1 & 0 & 0 & 0 & 0 & 1 & 0
\end{bmatrix} \quad
\boldsymbol{\Pi}_4 =
\begin{bmatrix}
	0 & 0 & 0 & 0 & 0 & 1 & 0 & 0 & 0 & 0\\
	0 & 0 & 0 & 0 & 0 & 0 & 1 & 0 & 0 & 0\\
	0 & 0 & 0 & 0 & 0 & 0 & 0 & 1 & 0 & 0\\
	0 & 0 & 0 & 0 & 0 & 0 & 0 & 0 & 1 & 0\\
	0 & 0 & 0 & 0 & 0 & 0 & 0 & 0 & 0 & 1
\end{bmatrix}
$$

$$\boldsymbol{\Pi}_2 = 
\begin{bmatrix}
	0 & 0 & 0 & 0 & 0 & 1 & 0 & 0 & 0 & 0\\
	0 & 0 & 0 & 0 & 1 & 0 & 0 & 0 & 0 & 1\\
	0 & 0 & 0 & 1 & 0 & 0 & 0 & 0 & 1 & 0\\
	0 & 0 & 1 & 0 & 0 & 0 & 0 & 1 & 0 & 0\\
	0 & 1 & 0 & 0 & 0 & 0 & 1 & 0 & 0 & 0\\
	1 & 0 & 0 & 0 & 0 & 1 & 0 & 0 & 0 & 0\\
	0 & 0 & 0 & 0 & 1 & 0 & 0 & 0 & 0 & 0\\
	0 & 0 & 0 & 1 & 0 & 0 & 0 & 0 & 0 & 1\\
	0 & 0 & 1 & 0 & 0 & 0 & 0 & 0 & 1 & 0
\end{bmatrix} \quad
\boldsymbol{\Pi}_5 =
\begin{bmatrix}
	1 & 0 & 0 & 0 & 0 & 0 & 0 & 0 & 0 & 0\\
	0 & 1 & 0 & 0 & 0 & 0 & 0 & 0 & 0 & 0\\
	0 & 0 & 1 & 0 & 0 & 0 & 0 & 0 & 0 & 0\\
	0 & 0 & 0 & 1 & 0 & 0 & 0 & 0 & 0 & 0\\
	0 & 0 & 0 & 0 & 1 & 0 & 0 & 0 & 0 & 0
\end{bmatrix}
$$
%\nocite{Wagh08}

\bibliographystyle{IEEEtran}
\bibliography{IEEEabrv,rs}

\end{document}